\documentstyle[preprint,epsfig,aps]{revtex}
\begin{document}
\draft
 
\title{ S-wave $\eta^{\prime}$-proton FSI; phenomenological analysis 
        of near-threshold production of $\pi^{0}$, $\eta$, and $\eta^{\prime}$
        mesons in proton-proton collisions.  }        

\author{P.~Moskal$^{1,2}$,
        H.-H.~Adam$^3$,
        J.T.~Balewski$^{4,5}$,
        A.~Budzanowski$^4$,
        J.~Budzi{\'n}ski$^{1}$,
        D.~Grzonka$^2$,
        L.~Jarczyk$^1$,
        A.~Khoukaz$^3$,
        K.~Kilian$^2$,
        P.~Kowina$^{2,6}$,
        N.~Lang$^3$,
        T.~Lister$^3$,
        W.~Oelert$^2$,
        C.~Quentmeier$^3$,
        R.~Santo$^3$,
        G.~Schepers$^{2,3}$,
        T.~Sefzick$^2$,
        S.~Sewerin$^2$,
        M.~Siemaszko$^6$,
        J.~Smyrski$^1$,
        A.~Strza{\l}kowski$^1$,
        M.~Wolke$^2$,
        P.~W{\"u}stner$^7$,
        W.~Zipper$^6$
       }
\address{$^1$ Institute of Physics, Jagellonian University, PL-30-059 Cracow, Poland}
\address{$^2$ IKP, Forschungszentrum J\"{u}lich, D-52425 J\"{u}lich, Germany}
\address{$^3$ IKP, Westf\"{a}lische Wilhelms--Universit\"{a}t, D-48149 M\"unster, Germany}
\address{$^4$ Institute of Nuclear Physics, PL-31-342 Cracow, Poland}
\address{$^5$ IUCF, Bloomington, Indiana, IN-47405, USA}
\address{$^6$ Institute of Physics, University of Silesia, PL-40-007 Katowice, Poland}
\address{$^7$ ZEL,  Forschungszentrum J\"{u}lich, D-52425 J\"{u}lich,  Germany}
 
\date{\today}
\maketitle
\begin{abstract}
     We describe a novel technique for comparing total cross sections for the 
     reactions $pp\rightarrow pp\pi^{0}$, $pp\rightarrow pp\eta$, and 
     $pp\rightarrow pp\eta^{\prime}$ close to threshold. 
     The initial and final  state proton-proton interactions are factored out
     of the total cross section, and the dependence of this reduced 
     cross section on the volume of phase space is discussed.  
     Different models of the  proton-proton interaction are compared. 
     We argue that the scattering length of the S-wave $\eta^{\prime}$-proton 
     interaction is of the order of 0.1~fm.
\end{abstract}


New results on $\eta$ and $\eta^{\prime}$ meson production in the   reaction
$pp \rightarrow ppX$, measured very recently at the COSY-11 
facility~\cite{smyrskipl,moskalpl}, together with previous 
data~\cite{hiboupl,moskalprl,calenpdeta,caleneta,pinot,bergdolt},  
determine the energy dependence of the near-threshold total cross section
with a precision comparable to the measurements of the reaction 
$pp\rightarrow pp\pi^{0}$~\cite{meyer92,bondar95}.
These new data encouraged us to perform a phenomenological analysis similar 
to those of references~\cite{sibifewbody,machner,bernard}. 
Here we  concentrate on $\pi^{0}$,  $\eta$, and $\eta^{\prime}$ meson production, 
and complete the analysis of these references by taking into account the 
interaction between the incident protons, and by introducing a new representation 
of the data. The production rates of $\pi^{0}$,  $\eta$, and $\eta^{\prime}$ mesons 
will be compared as a function of the available phase space.   
We will study the phase-space dependence of the quantity $|M_{0}|$ which is derived
from the total cross section with the ISI and pp-FSI factored out.  
Consideration of the dependence of $|M_{0}|$ on phase space allows us to infer 
the $\eta$-proton and $\eta^{\prime}$-proton interactions. To avoid large ambiguities 
due to differences in pp-FSI models, we normalize the $|M_{0}|$ for 
$\eta$ and $\eta^{\prime}$ mesons to the one for the $\pi^{0}$ meson and show that the resulting
$\eta^{\prime}$-proton interaction is comparable to the $\pi^{0}$-proton one.

In general, the total cross section is presented as a function of the dimensionless 
parameter~$\eta_{M}$~\cite{meyer92,bondar95,machner}~\footnote{
      In order to avoid ambiguities with the abbreviation for the eta-meson, we introduce an additional 
      suffix M  for this parameter, which usually is called  $\eta$.
      },
which is
defined as the maximum meson momentum in  units of meson 
mass~\begin{math}\begin{displaystyle}(\eta_{M} = \frac{q_{max}}{M})\end{displaystyle}\end{math},
or as a function of the center-of-mass excess energy~$Q$~\cite{hiboupl,caleneta,bergdolt},
where nonrelativistically these variables are related by:
\begin{equation}
\eta_{M}^{2} = \frac{4 \ m_{p}}{ 2 \ m_{p} \ M + M^{2}} \ Q,
\end{equation}
with $m_{p}$ and M denoting the proton and meson mass, respectively.
The above equation shows that the proportionality factor between the variables Q and $\eta_{M}^{2}$ changes
for different mesons, since the masses of the $\pi^{0}$, $\eta$, and $\eta^{\prime}$ are distinct.
Hence, depending which variable is selected, 
the relation between the cross section values changes for different mesons.
For example, as shown in reference~\cite{moskalprl}
the $\eta$ meson production-cross-section exceeds the $\pi^{0}$ cross section
by about a factor of five using $\eta_{M}$,
whereas the $\pi^{0}$ meson cross section is always larger when employing the $Q$ scale. 

The total cross section for $pp\rightarrow ppX$ is in general an integral over phase space,
weighted by the square of the transition matrix element and normalized to the 
incoming flux factor~F:
\begin{equation}
  \sigma_{pp\rightarrow ppX}=
  \frac{ 1 }{ F } \displaystyle \int dV_{ps} \ |M_{pp\rightarrow ppX}|^{2},
  \label{eqcrossform}
\end{equation}
where
 X stands for the $\pi^{0},\eta$ or $\eta^{\prime}$ meson,
 $V_{ps}$ denotes the phase space volume, and 
F~=~\begin{math}2~(2\pi)^{5}\sqrt{s~(s~-~4m_{p}^{2})}\end{math}~~\cite{byckling}, 
with s being the square of the total energy in the center-of-mass frame.
The transition matrix element for the reaction $pp\rightarrow ppX$,$~~$ 
$M_{pp\rightarrow ppX}$,$~~$ incorporates the production mechanism and both the 
initial (ISI) and final (FSI) state interactions.
In analogy with the {\em Watson-Migdal} approximation~\cite{wats52} for two body processes,
we may assume that the complete transition amplitude for a production process
$M_{pp\rightarrow ppX}$ factorizes approximately as~\cite{moalem1}:
\begin{equation}
   |M_{pp \rightarrow ppX}|^{2}   \approx  |M_{0}|^{2} \cdot |M_{FSI}|^{2} \cdot ISI,
\label{M0FSIISI}
\end{equation}
where  $M_{0}$ represents the total
production amplitude, $M_{FSI}$  describes the elastic interaction among particles 
in the exit channel, and ISI denotes the reduction factor 
due to the interaction of the colliding protons.

Equation~(\ref{eqcrossform})
suggests that a natural variable for comparing  the total cross sections for different
mesons may be the 
volume of available phase space,\footnote{ 
      In the nonrelativistic approximation, Q~$<<$~M (which is justified at 
      threshold), $V_{ps}$ is proportional
      to the fourth power of the variable $\eta_{M}$ or the square of the excess energy $Q$;
      \begin{math}\begin{displaystyle} V_{ps} = \int dV_{ps} = \eta_{M}^{4} \ \sqrt{M + 2\ m_{p}}
      \cdot m_{p}^{-1} \ M^{\frac{5}{2}} \ \pi^{3} \ 2^{-5} \end{displaystyle}\end{math}
     } \ $V_{ps} = \int dV_{ps}$.
Figure~\ref{cfisi_Vps} shows the yield of $\pi^{0}$, $\eta$, and 
$\eta^{\prime}$ mesons in the proton-proton interaction as a function of 
available phase space volume. The yield is defined as the cross section  
multiplied by the corresponding flux factor, and divided by the ISI factor.
Close to threshold the initial state interaction, which reduces the total cross section,
is dominated by proton-proton scattering in the $^{3}P_{0}$ state. 
As shown by Hanhart and~Nakayama~\cite{hannak}, this ISI may be estimated  in 
terms of  phase shifts and inelasticities. The ISI factor is close to unity for
pion production, and amounts to $\sim$~0.2~\cite{hannak} and 
$\sim$~0.33~\cite{nakayama} respectively for $\eta$ and $\eta^{\prime}$
mesons at threshold. As shown in reference~\cite{hannak}
the initial state interaction may be taken into account by multiplying the 
theoretical  cross section by these factors. 
This  justifies our choice of the dimensionless quantity 
\begin{math} \begin{displaystyle} \frac{\sigma \cdot F}{ISI} \end{displaystyle} \end{math},
which  depends only on the primary production amplitude~$M_{0}$
and on the final state interaction among the produced particles.
Numerically we have confirmed that within the present model 
the effect of the pp-FSI is approximately independent of the produced meson 
mass throughout the phase space volume under investigation.
In fact, the difference of the effect on the pp-FSI is about 1$\%$ for $\eta$ 
and $\eta^{\prime}$ production and it is only about 10$\%$ larger than for  the 
case of the $\pi^{0}$ production. This is in line with the G.~F\"aldt and 
C.~Wilkin model which predict that the pp-FSI depends on the excess energy 
only~\cite{faldtwilk}.

Thus, the observed differences in the production yield for the different mesons,
as presented in Figure~\ref{cfisi_Vps}, may be attributed directly to the 
square of the primary production amplitude $|M_{0}|^{2}$ convoluted with the 
FSI of the particles in the exit channel.
Comparing the cross sections with flux-factor and ISI corrections one can see
that over the relevant range of $V_{ps}$ the dynamics for $\eta^{\prime}$ meson 
production is about six times weaker than for the $\pi^{0}$ meson, 
which again is a further factor of six  weaker than that of the $\eta$ meson.
 
Employing equations~(\ref{eqcrossform}) and~(\ref{M0FSIISI}) and  given the 
two additional assumptions that in the exit channel only the proton-proton 
interaction is significant 
(\begin{math} M_{FSI} = M_{pp\rightarrow pp} \end{math}),
and that the primary production amplitude is constant over the studied range of phase space,
it is possible  to calculate the quantity $|M_{0}|$. 
From this point we no longer refer to $M_{0}$ as the
primary production amplitude, because the assumption (\begin{math} M_{FSI} = M_{pp\rightarrow pp} \end{math})
implicitly  shifts  the proton-meson FSI from $|M_{FSI}|$ to $|M_{0}|$.

To evaluate  $|M_{0}|$ we considered
three possible parametrizations of the pp-FSI enhancement factor $|M_{pp\rightarrow pp}|^{2}$,
which are presented in Figure~\ref{Mpppp}.
The solid line shows the squared proton-proton amplitude~\cite{morton}: 
\begin{equation}
 M_{pp\rightarrow pp} = \frac{e^{\delta_{pp}({^{1}S_{0}})} \cdot \sin{\delta_{pp}({^{1}S_{0}})} }{C \cdot k},
\label{amppp}
\end{equation}
where \begin{math} C^{2} = \frac{2\pi\eta_{c}}{e^{2\pi\eta_{c}} -1} \end{math} 
is the Coulomb penetration factor~\cite{bethe}, 
$\eta_{c}$  is the relativistic  Coulomb parameter $\eta_{c}=\alpha / v$,
with $\alpha$~-~the fine structure constant and $v$~-~the  proton velocity
in the rest frame of the other proton.
The phase-shifts \begin{math} \delta_{pp}({^{1}S_{0}}) \end{math} are
calculated according to the modified Cini-Fubini-Stanghellini
formula including the Wong-Noyes Coulomb correction~\cite{naisse,noyeslip,noyes},
\begin{eqnarray}
 C^{2} \ p \  ctg\delta_{pp}+2 \ p \ \eta_{c} \ h(\eta_{c})=-\frac{1}{a_{pp}} + 
  \frac{b_{pp}\ p^{2}}{2}
  - \frac{P_{pp} \ p^{4}}{1 + Q_{pp} \ p^{2}},
  \label{CFS}
\end{eqnarray}
where $h(\eta_{c}) = -ln(\eta_{c}) -0.57721 + \eta_{c}^{2} \
\sum_{n=1}^{\infty}\frac{1}{n\cdot (n^{2}+\eta_{c}^{2})}$~\cite{jack50}.
The phenomenological quantities $a_{pp}=-7.83$~fm
and $b_{pp}=2.8$~fm  denote the scattering length and effective range~\cite{naisse},
respectively. The parameters $P_{pp}=0.73$~fm$^{3}$ and $Q_{pp}=3.35$~fm$^{2}$ 
are related to the detailed shape of the nuclear 
potential and derived  from a one-pion-exchange model~\cite{naisse}.
These  calculations give values which are in a good agreement with the
phase shifts of the VPI partial wave analysis~\cite{arndt}, shown
as solid circles, and with the phase shifts of the Nijmegen analysis~\cite{nijmpsa}, shown as open squares. 

The dashed line presents the enhancement from the
proton-proton interaction, $|M_{pp\rightarrow pp}|^{2}$,
estimated as an inverse of the squared Jost function,
with the Coulomb interaction  included~\cite{druzhinin}.
In this case $|M_{pp\rightarrow pp}|^{2}$ is a dimensionless factor which
approaches zero as the relative
proton momentum k$\rightarrow$0, peaks sharply at k~$\approx$~25~MeV/c, and
asymptotically approaches unity for large relative proton-proton momentum.
The solid and dashed lines agree quite well for small relative protons momentum.

The dotted line corresponds to the inverse   of the squared Jost function,
calculated using the formulas of references~\cite{shyammosel,watson}, 
also corrected for the Coulomb force.
The presented prescriptions evidently differ significantly, 
especially for k larger than $\approx$~50~MeV/c.
 
The curves in  Figure~\ref{Mpppp} are arbitrarily normalized to the same maximum value 
as found in reference~\cite{druzhinin}.
Note that the values of $|M_{0}|$ extracted will depend on the 
absolute values of the 
enhancement factor
$|M_{pp\rightarrow pp}|^{2}$, which is not well established. For example, the formula given by 
Druzhinin et al.~\cite{druzhinin} 
leads to a  maximum value of $|M_{pp\rightarrow pp}|^{2}$ of about 50, 
whereas the authors of reference~\cite{sibifewbody} find a value of about 20. 
Due to this wide variation, in the following we will discuss only the relative 
phase space dependence of $|M_{0}|$, rather than the absolute magnitude.
 
Figure~\ref{M0cfs} compares values extracted for $|M_{0}|$
near  threshold production of $\pi^{0}$, $\eta$, and $\eta^{\prime}$  mesons, 
using $M_{pp\rightarrow pp}$ as shown in Figure~\ref{Mpppp} by the solid line. 
$|M_{0}|$ is arbitrarily normalized to unity for large $V_{ps}$, separately for each meson.
If the assumptions used in the derivation of $|M_{0}|$ were strictly satisfied
the values obtained would be equal to one as depicted by the solid line.
It can be seen, however, that in the case of the $\eta$ meson, $|M_{0}|$ grows
with decreasing phase space volume. The observed deviation from unity is too large
to be plausibly assigned to a variation in the primary production amplitude;
the calculations of Moalem et al.~\cite{moalem} show that 
the primary production amplitude should change only  by a few per cent 
in this energy range.
Therefore, the observed behaviour of $|M_{0}|$ may be attributed to an attractive $\eta$-proton interaction,
which was neglected in its derivation.
The results for the $\pi^{0}$ and $\eta^{\prime}$ production,
   apart from the points closest to threshold\footnote{
   Due to the steep decrease of the total cross section near threshold,
   a small change in the energy (0.2~MeV) lifts the points to significantly higher values.
   Moreover, at very low energies, nuclear and Coulomb scattering are expected to compete.
   The limit is at approximately 0.8~MeV  proton energy in the rest frame of the other proton,
   where the Coulomb penetration factor~C$^{2}$ is equal to 0.5~\cite{jack50}.
   Thus, careful analysis is required at small excess energies,
   where the Coulomb interaction dominates.
  },
show that $|M_{0}|$ is indeed effectively constant over the region of the phase space studied,
indicating that both the $\pi^{0}$-proton and $\eta^{\prime}$-proton
FSIs are too weak to be observable 
at current experimental
accuracy. In the case of the $\pi^{0}$ this result  was expected, since the S-wave $\pi$-proton
interaction is much weaker than the $\eta$-proton one. 
The  real part of the scattering
length, a$_{p\pi}$~=~0.13~fm~\cite{sigg}, is about six times 
smaller than a$_{p\eta}$~=~0.75~fm~\cite{greenwycech}. 
Regarding the $\eta^{\prime}$ meson,
there is no experimental evidence for the $\eta^{\prime}$-proton interaction except 
for a very conservative upper limit on the real value 
of the $\eta^{\prime}$-proton scattering length ~$|Re(a_{\eta^{\prime}p})|$~$<$~0.8~fm,
as estimated in reference~\cite{moskalpl}. 
In deriving this $\eta^{\prime}$-proton estimate
the dashed line in Figure~\ref{Mpppp} was assumed for the proton-proton FSI.

 Different prescriptions for the proton-proton enhancement factors will obviously
give different results for $|M_{0}|$.
Figure~\ref{M0nisk} for example shows $|M_{0}|$ 
for the various meson production channels,
as extracted when using the distribution of  $|M_{pp\rightarrow pp}|^{2}$ 
depicted in Figure~\ref{Mpppp} as the dotted line. 
The $|M_{0}|$ behaviour for  $\pi^{0}$ and $\eta^{\prime}$ mesons
is now qualitatively different from that shown in Figure~\ref{M0cfs},
and would indicate an unreasonably strong attractive interaction between
these mesons and the proton. 
Results for $|M_{0}|$ obtained with $|M_{pp\rightarrow pp}|^{2}$
as described by  
Druzhinin et al.~\cite{druzhinin}~(dashed line in Figure~\ref{Mpppp}),
which  is close to that presented in Figure~\ref{M0cfs},
can be found in~\cite{moskalstori}.
 
To minimize ambiguities that result from  uncertainties 
in the proton-proton scattering amplitude, we consider
the ratio $|M^{\eta(\eta^{\prime})}_{0}|/|M^{\pi^{0}}_{0}|$ . 
At first order the integral of $|M_{pp\rightarrow pp}|^{2}$ 
over phase space  is independent of the meson produced. 
The transition amplitude for $\eta$ and $\eta^{\prime}$ production $|M^{\eta(\eta^{\prime})}_{0}|$
is therefore normalized to the one for the $\pi^{0}$ production $|M^{\pi^{0}}_{0}|$;
this should be independent of the model used for the determination of $|M_{pp\rightarrow pp}|^{2}$, 
and will allow an estimate of the relative strength of the $\pi^{0}$-proton and 
$\eta(\eta^{\prime})$-proton interactions.
Indeed, we found that within errors the ratio $|M^{\eta(\eta^{\prime})}_{0}|/|M^{\pi^{0}}_{0}|$ 
does not depend on the model used for  $|M_{pp\rightarrow pp}|^{2}$.
As an example, in Figure~\ref{mmnisk} we show this ratio as obtained
from the  amplitude $|M_{pp\rightarrow pp}|^{2}$
presented as the dotted line in Figure~\ref{Mpppp}. Figure~\ref{mmnisk}a shows an increasing strength
of $|M_{0}|$  for the $\eta$ production at low $V_{ps}$,
indicating a strong $\eta$-proton FSI, as was discussed previously for the cross section ratio
by Cal\'{e}n et al.~\cite{caleneta}.
Note also that  the ratio for the $\eta^{\prime}$ meson 
is constant over the phase space range considered (Figure~\ref{mmnisk}b).
This observation, and the fact that theoretical
calculations predict that the primary production amplitude is constant 
to within a few per cent~\cite{nakayama,gedalin}
independent of the mechanism assumed, allows us to conclude that
the $\eta^{\prime}$-proton scattering parameters are comparable to the $\pi^{0}$-proton ones. 
The $\eta^{\prime}$-proton scattering length is therefore about 
0.1~fm, similar to  the $\pi^{0}$-proton scattering length.
\newpage
{\bf{Acknowledgements}}\\
One of the authors (P.M.)
acknowledges financial support from the Forschungszentrum
J\"ulich and the Foundation for Polish Science.
We would like to thank T. Barnes for his careful reading of the manuscript.
This research project was supported in part by
the BMBF (06MS881I),
the Bilateral Cooperation between Germany and Poland 
represented by the Internationales B\"{u}ro DLR for the BMBF (PL-N-108-95) 
and by the Komitet Bada{\'n} Naukowych KBN, 
and by FFE grants (41266606 and 41266654)
from the Forschungszentrum J\"{u}lich.

\newlength{\ts}
\setlength{\ts}{\textwidth}
\addtolength{\ts}{-\columnsep}
\setlength{\ts}{0.5\ts}

\begin{figure}[h]
\epsfxsize=\ts
\vspace{2.5cm}
\centerline{\epsfig{figure=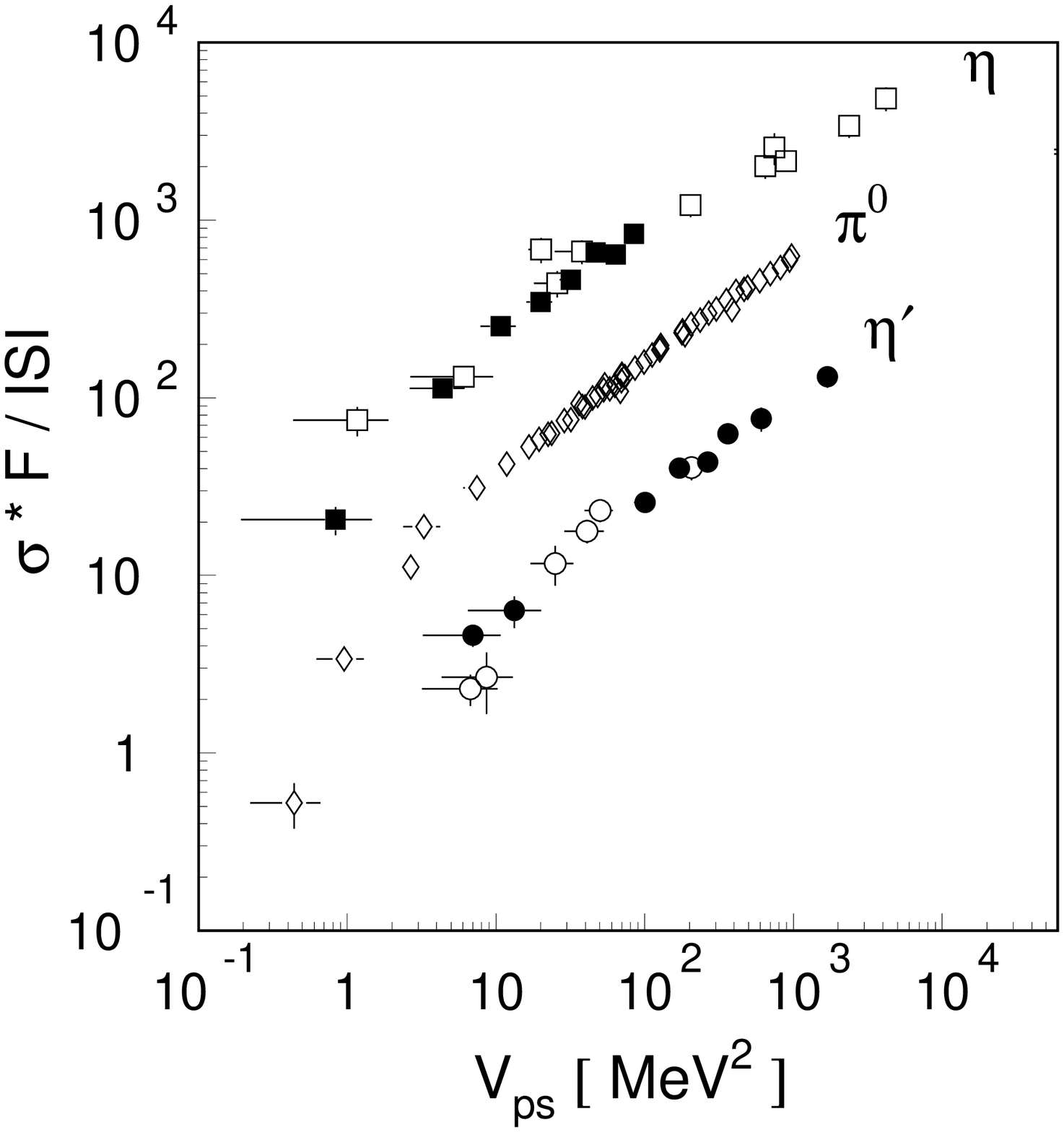,height=11.0cm,angle=0}}
\vspace{2.0cm}
\caption{ Total cross section multiplied by the flux factor F and 
          divided by the initial-state-iteraction reduction factor ISI
          versus the available phase space volume for the reactions
          $pp\rightarrow pp\eta$~(squares~\protect\cite{smyrskipl,hiboupl,caleneta,pinot,bergdolt}), \ \ \ \ 
          $pp\rightarrow pp\pi^{0}$~(diamonds~\protect\cite{meyer92,bondar95,meyer90}), \ \ \ \ 
          and $pp\rightarrow pp\eta^{\prime}$~(circles~\protect\cite{moskalpl,hiboupl,moskalprl}).
          The filled symbols are recent COSY~-~11 results~\protect\cite{smyrskipl,moskalpl}.
          Note that an increase in the excess energy from Q~=~0.5~MeV
          to Q~=~30~MeV corresponds a growth of  phase space volume
          by about three orders of magnitude.
        }
\label{cfisi_Vps}
\end{figure}

\newpage

\begin{figure}[h]
\epsfxsize=\ts
\vspace{2.5cm}
\centerline{\epsfig{figure=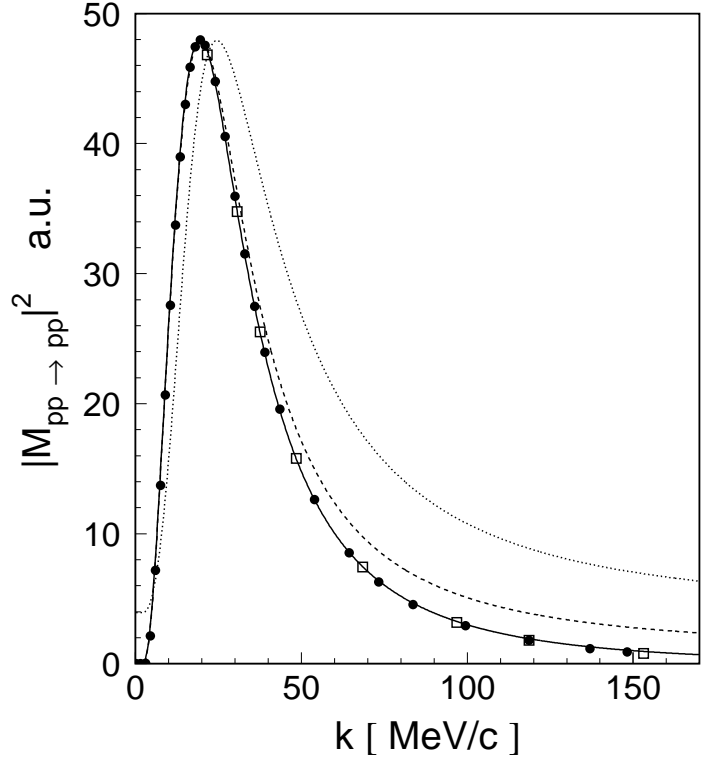,height=11.0cm,angle=0}}
\vspace{2.0cm}
\caption{ Square of  the proton-proton scattering amplitude
            versus k, the proton momentum  in the proton-proton subsystem.
            These are references \protect\cite{morton,naisse}~(solid line),
            \protect\cite{druzhinin}~(dashed line) and
            \protect\cite{shyammosel,watson}~(dotted line). 
            The filled circles are extracted from~\protect\cite{arndt}, and
            the opened squares are from~\protect\cite{nijmpsa}.
            The curves and symbols are arbitrarily normalized  to be equal at maximum to the 
            result from reference~\protect\cite{druzhinin}, shown as the dashed line. \protect \\
          The presented range of momentum k  covers the allowed proton
          momenta in the excess energy range  Q~$<$~30~MeV.
        }
\label{Mpppp}
\end{figure}

\newpage

\begin{figure}[h]
 \unitlength 1.0cm
  \begin{picture}(12.2,18.0)
    \put(3.0,0.0){
       \epsfig{figure=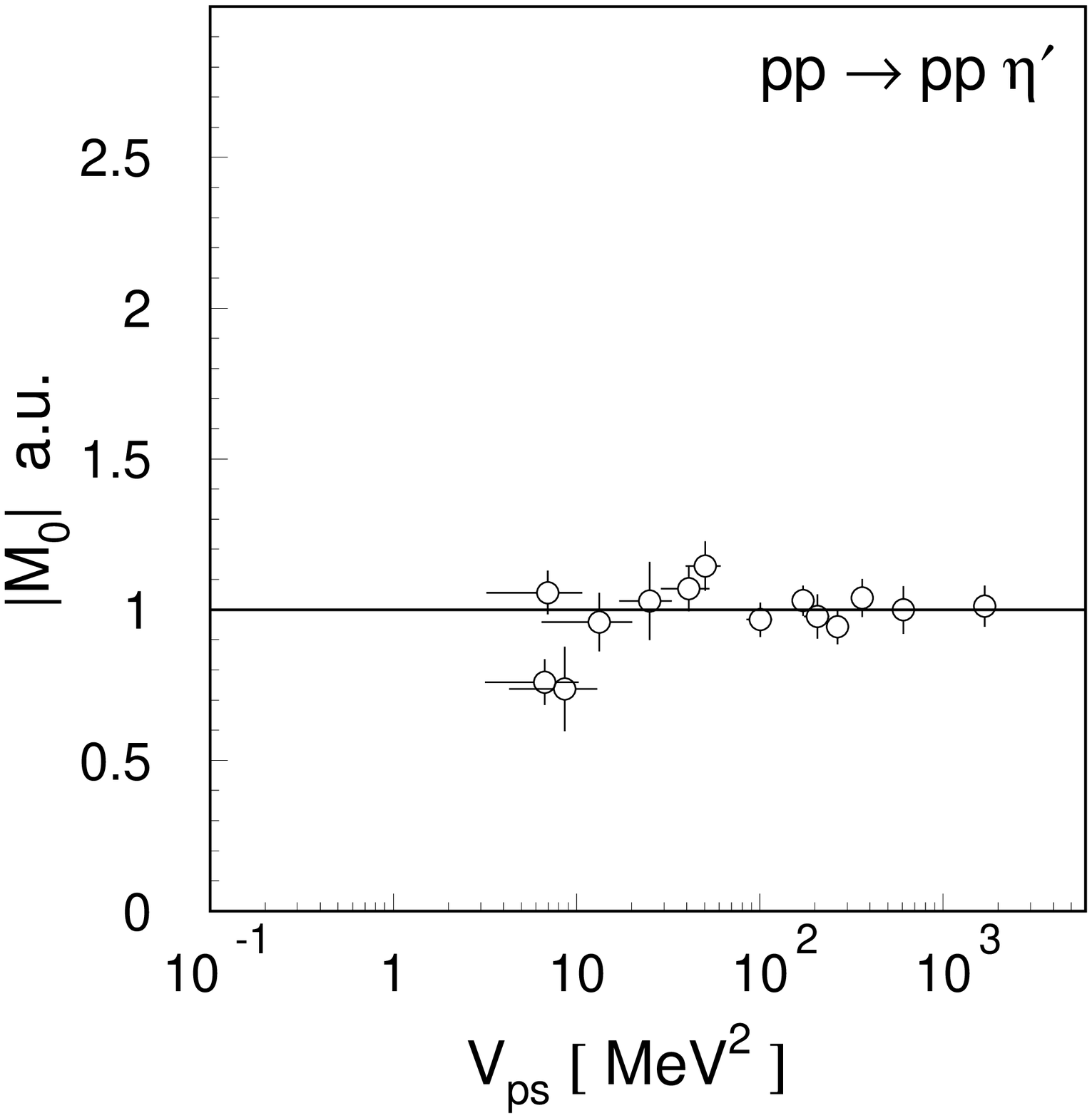,height=5.8cm,width=7.0cm,angle=0}
    }
       \put(9.5,1.5){
          { c)}
       }
    \put(3.0,4.5){
       \epsfig{figure=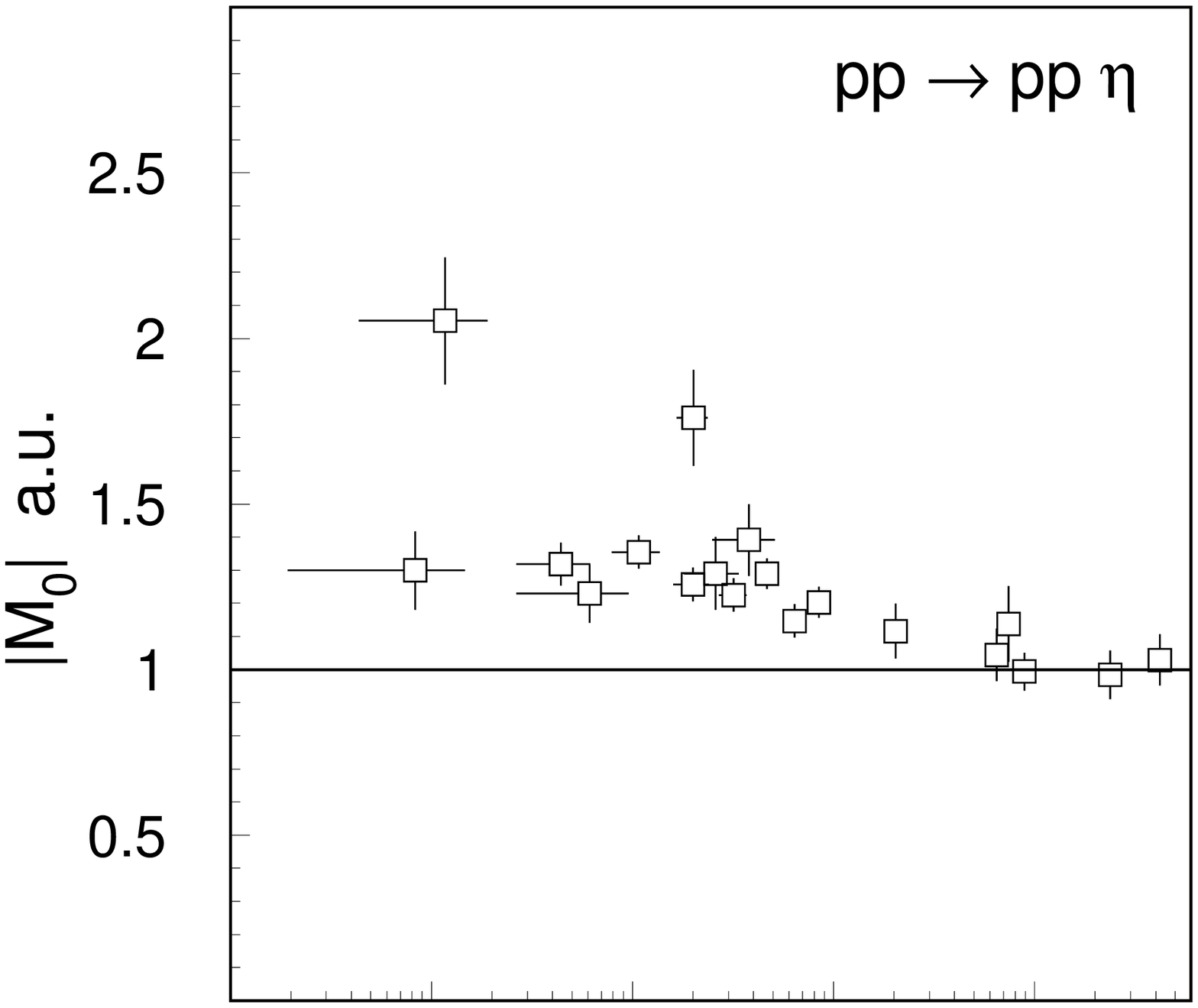,height=5.8cm,width=7.0cm,angle=0}
    }
       \put(9.5,6.0){
          { b)}
       }
    \put(3.0,9.0){
       \epsfig{figure=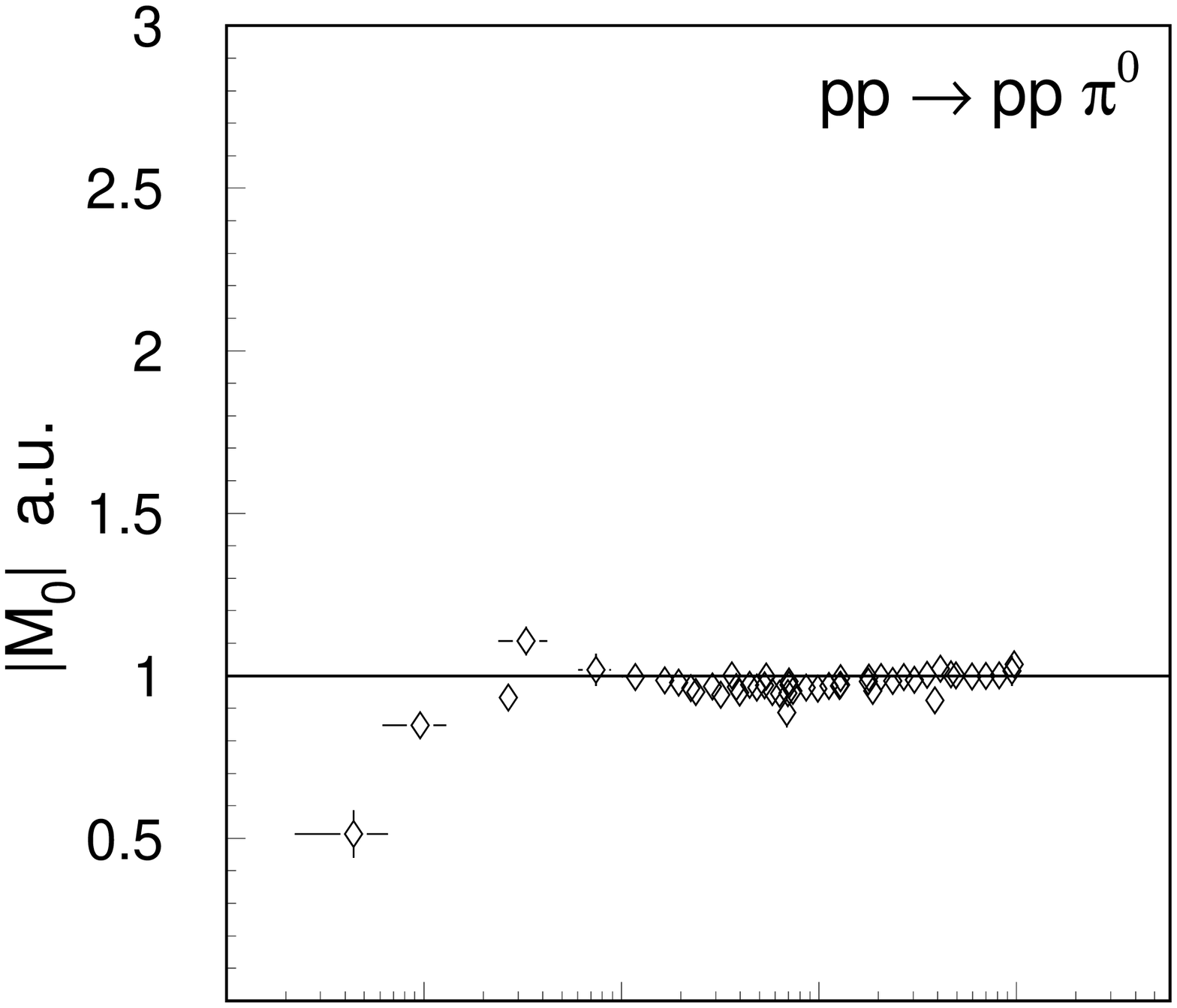,height=5.8cm,width=7.0cm,angle=0}
    }
       \put(9.5,10.5){
          { a)}
       }
  \end{picture}
  \vspace{1.4cm}
  \caption{
           The quantity $|M_{0}|$,
             extracted from the data of references
            \protect\cite{smyrskipl,moskalpl,hiboupl,moskalprl,caleneta,pinot,bergdolt,meyer92,bondar95,meyer90}. 
             The proton-proton scattering amplitude was
              calculated according to equation~(\protect\ref{amppp}), and the phase shifts were computed using
              equation~(\protect\ref{CFS}) (see the solid line in
	      Figure~\protect\ref{Mpppp}). 
        }
\label{M0cfs}
\end{figure}

\newpage

\begin{figure}[h]
 \unitlength 1.0cm
  \begin{picture}(12.2,18.0)
    \put(3.0,0.0){
       \epsfig{figure=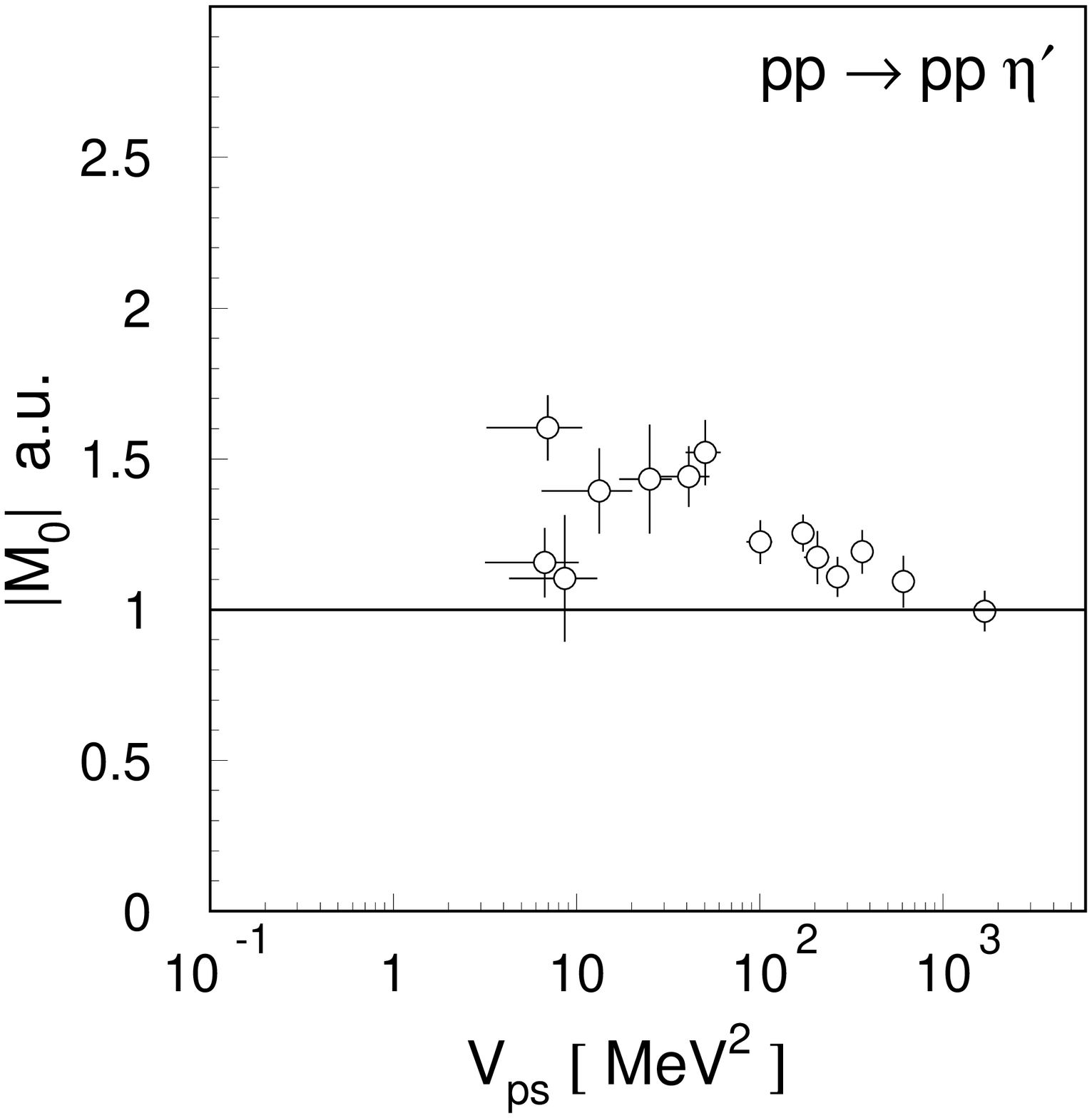,height=5.8cm,width=7.0cm,angle=0}
    }
       \put(9.5,1.5){
          { c)}
       }
    \put(3.0,4.5){
       \epsfig{figure=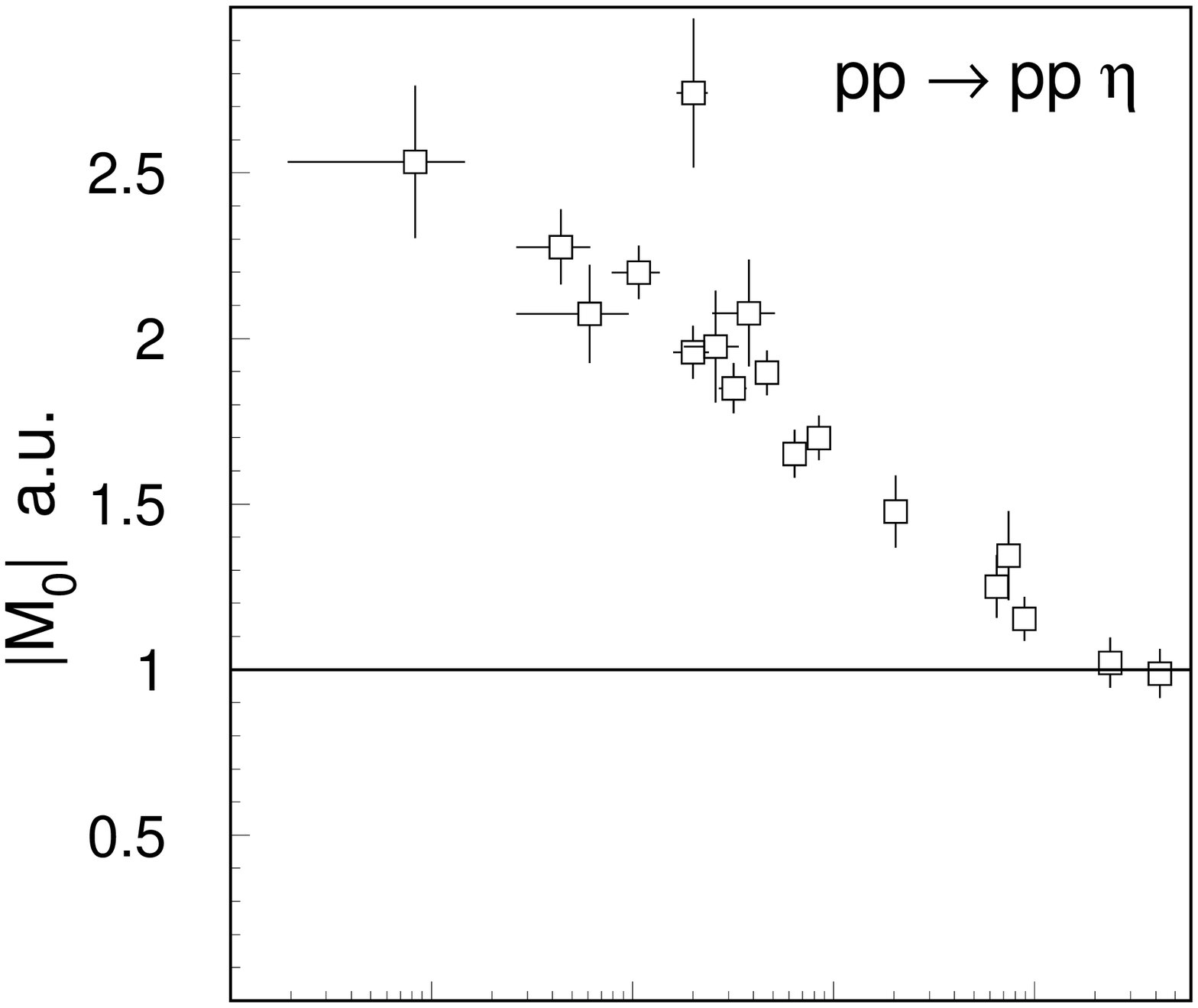,height=5.8cm,width=7.0cm,angle=0}
    }
       \put(9.5,6.0){
          { b)}
       }
    \put(3.0,9.0){
       \epsfig{figure=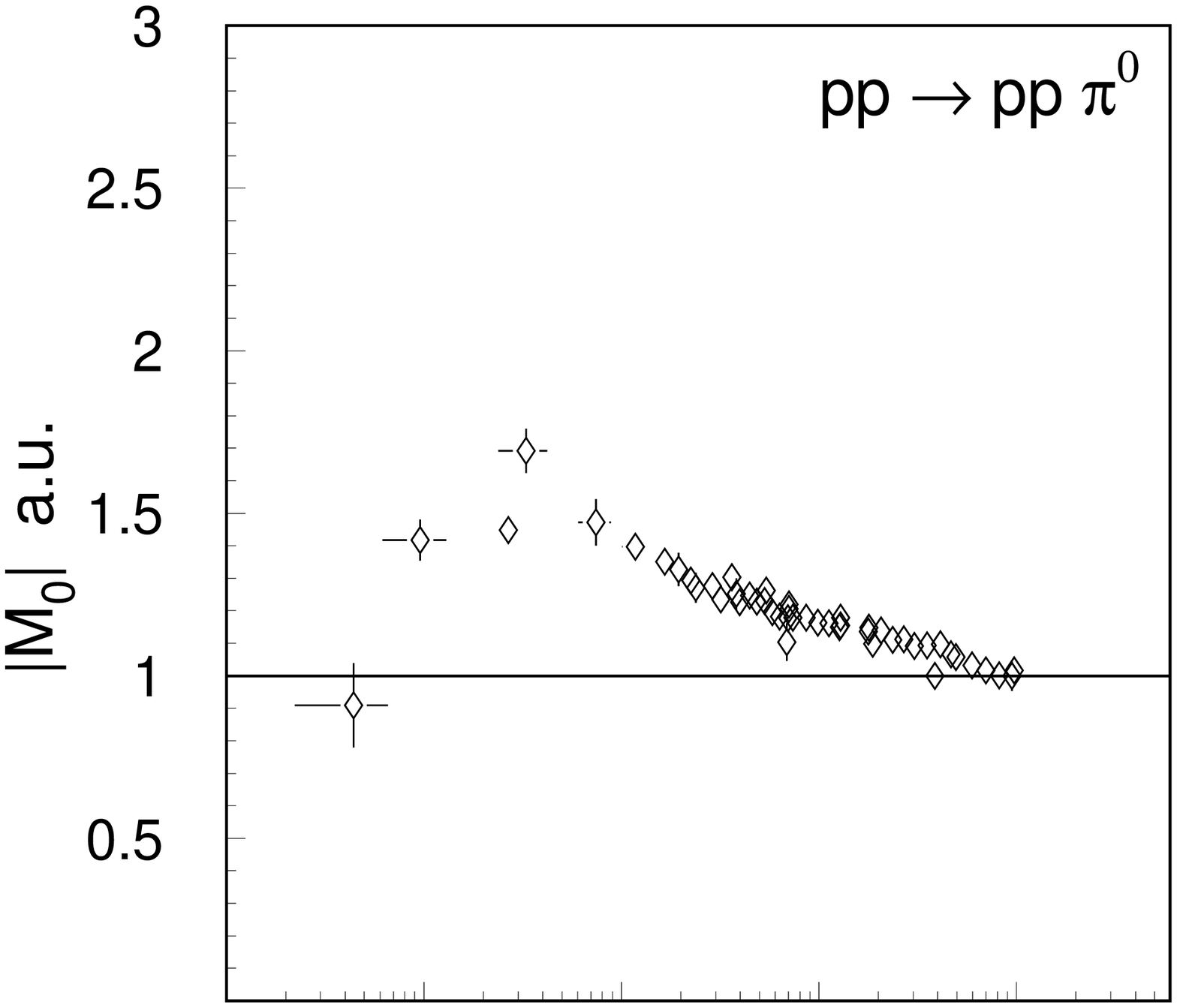,height=5.8cm,width=7.0cm,angle=0}
    }
       \put(9.5,10.5){
          { a)}
       }
  \end{picture}
  \vspace{1.4cm}
  \caption{ 
           The quantity $|M_{0}|$,
             extracted from the data of references
            \protect\cite{smyrskipl,moskalpl,hiboupl,moskalprl,caleneta,pinot,bergdolt,meyer92,bondar95,meyer90}. 
              The  enhancement due to  the pp-FSI was approximated
              by the inverse of the squared Jost function of references~\protect\cite{shyammosel,watson}
              (dotted line in Figure~\protect\ref{Mpppp}).
        }
\label{M0nisk}
\end{figure}

\newpage

\begin{figure}[h]
 \unitlength 1.0cm
  \begin{picture}(12.2,18.0)
    \put(3.0,0.0){
       \epsfig{figure=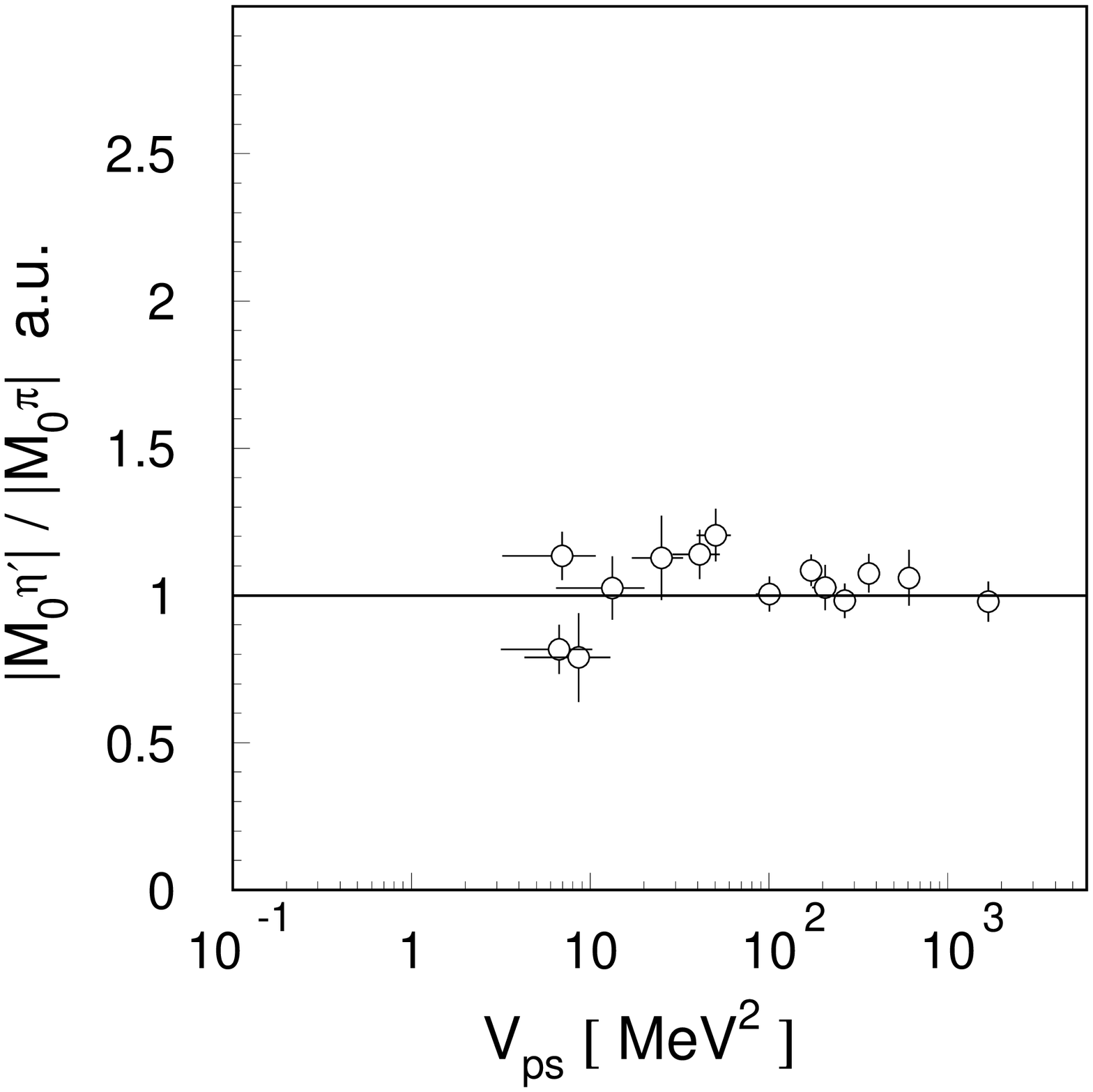,height=5.8cm,width=7.0cm,angle=0}
    }
       \put(9.5,1.5){
          { b)}
       }
    \put(3.0,4.5){
       \epsfig{figure=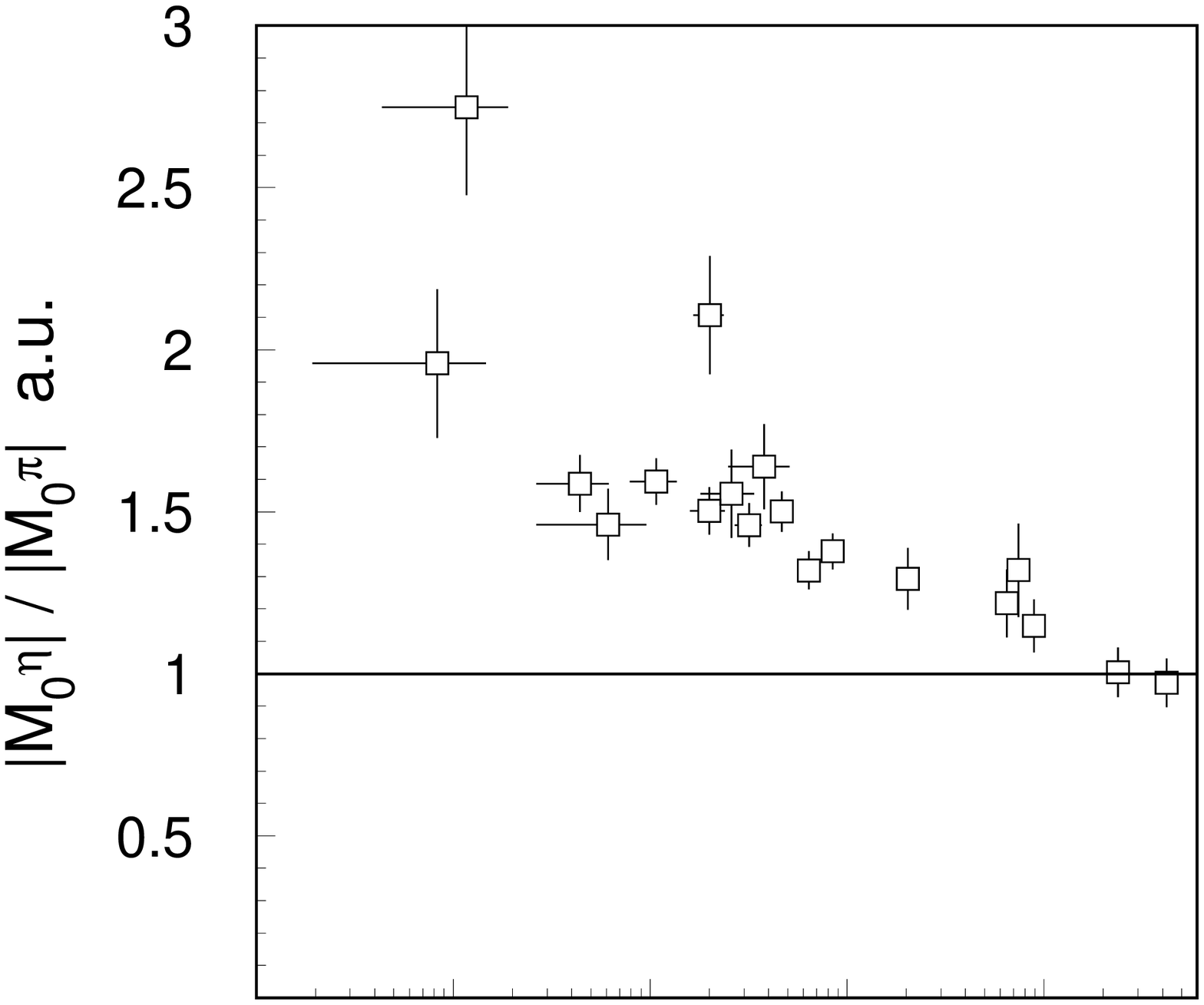,height=5.8cm,width=7.0cm,angle=0}
    }
       \put(9.5,6.0){
          { a)}
       }
  \end{picture}
  \vspace{1.4cm}
  \caption{
             The ratios of $~~$ a) $|M^{\eta}_{0}|/|M^{\pi^{0}}_{0}|$ 
             $~~$ and $~~$ b) $|M^{\eta^{\prime}}_{0}|/|M^{\pi^{0}}_{0}|$
             extracted from the data, assuming the
             pp-FSI enhancement factor depicted by the dotted line in 
	     Figure~\protect\ref{Mpppp}.
             $|M^{\pi^{0}}_{0}|$ 
             was calculated by interpolating  the points of 
	     Figure~\protect\ref{M0nisk}a.
        }
\label{mmnisk}
\end{figure}


\begin{thebibliography}{bib99} 
\bibitem{smyrskipl}
  J. Smyrski et al.,  Phys. Lett. {\bf B 474}  (2000) 182.
\bibitem{moskalpl}
  P. Moskal et al.,   Phys. Lett. {\bf B 474}  (2000) 416.
\bibitem{hiboupl}
  F. Hibou et al., Phys. Lett. {\bf B 438} (1998), 41.
\bibitem{moskalprl}
  P. Moskal et al.,   Phys. Rev. Lett. {\bf 80} (1998) 3202.
\bibitem{calenpdeta}
  H. Cal\'{e}n et al.,  Phys. Rev. Lett. {\bf 79} (1997), 2642. 
\bibitem{caleneta}
  H. Cal\'{e}n et al.,  Phys. Lett. {\bf B 366} (1996), 39.
\bibitem{pinot}
  E. Chiavassa  et al.,  Phys. Lett. {\bf B 322} (1994), 270.
\bibitem{bergdolt}
  A. M. Bergdolt et al., Phys. Rev. {\bf D 48} (1993), R2969.
\bibitem{meyer92}
  H. O. Meyer et al., Nucl. Phys. {\bf A 539} (1992) 633.
\bibitem{bondar95}
  A. Bondar et al.,   Phys. Lett. {\bf B 356} (1995) 8.
\bibitem{sibifewbody}
 A.~Sibirtsev and W.~Cassing, Few-Body Systems Suppl.\ {\bf 99} (1999) 1.
\bibitem{machner}
  H. Machner, J. Haidenbauer, J. Phys. {\bf G 25} (1999) R231.
\bibitem{bernard}
  V.~Bernard, N.~Kaiser, Ulf-G.~Mei{\ss}ner,  Eur. Phys. J. {\bf A 4} (1999), 259.
\bibitem{byckling}
 E. Byckling, K. Kajantie, \
  {\em Particle Kinematics}, \\
  John Wiley $\&$ Sons  Ltd.  (1973) 
\bibitem{hannak}
  C. Hanhart, K. Nakayama, Phys. Lett. {\bf B 454} (1999) 176.
\bibitem{nakayama}
  K.~Nakayama et al., Phys. Rev. {\bf C 61} (2000) 024001.
\bibitem{faldtwilk}
  G. F\"aldt and C. Wilkin, Phys. Lett. {\bf B 382} (1996), 209.
\bibitem{wats52}
   K.M. Watson,
   Phys. Rev. {\bf 88} (1952), 1163
\bibitem{moalem1}
  A. Moalem  et al.,  Nucl. Phys. {\bf A 589} (1995) 649.
\bibitem{morton}
  B.~J.~Morton et al., Phys. Rev. {\bf 169} (1968) 825.
\bibitem{bethe}
  H. A. Bethe, Phys. Rev. {\bf 76} (1949) 38.
\bibitem{naisse}
  J. P. Naisse,  Nucl. Phys. {\bf A 278} (1977), 506.
\bibitem{noyeslip}
  H.~P.~Noyes, H.~M.~Lipinski, Phys. Rev. {\bf C 4} (1971), 995.
\bibitem{noyes}
  H.~P.~Noyes, Ann. Rev. Nucl. Sci. {\bf 22} (1972), 465. 
\bibitem{jack50}
   J.D. Jackson, J.M. Blatt, \
   Rev. of Mod. Phys. {\bf 22} (1950), 77
\bibitem{arndt}
  R.A. Arndt et al., Phys. Rev. {\bf C 56} (1997) 3005. \\
  The Virginia Tech Partial-Wave Analysis Facility (SAID)\\
  http://said.phys.vt.edu/said$\_$branch.html
\bibitem{nijmpsa}
  V.G.J. Stoks et al., Phys. Rev. {\bf C 48} (1993) 792.
\bibitem{druzhinin}
  B. L. Druzhinin, A. E. Kudryavtsev, V. E. Tarasov,
    Z.~Phys. {\bf A 359} (1997), 205.
\bibitem{shyammosel}
 R. Shyam, U. Mosel, Phys. Lett. {\bf B 426} (1998) 1.
\bibitem{watson}
  M.L. Goldberger, K.M. Watson, Collision Theory, Wiley, New York, 1964.
\bibitem{moalem}
  A. Moalem  et al.,  Nucl. Phys. {\bf A 600} (1996), 445.
\bibitem{sigg}
 D. Sigg et al., Nucl. Phys. {\bf A 609} (1996) 269, and Nucl. Phys. {\bf A 617} (1997) 526
\bibitem{greenwycech}
  A. M. Green, S. Wycech,  Phys. Rev. {\bf C 55} (1997), R2167.
\bibitem{moskalstori}
    P. Moskal et al., 
            4th International Conference on Physics at Storage Rings,
           Bloomington, Indiana, USA, 1999, in press by  American Institute of Physics.\\
    http://ikpe1101.ikp.kfa-juelich.de/
\bibitem{gedalin}
  E.~Gedalin, A.~Moalem, L.~Razdolskaja, Nucl. Phys. {\bf A 650} (1999), 471. 
\bibitem{meyer90}
  H. O. Meyer et al.,  Phys. Rev. Lett. {\bf 65} (1990), 2846.
 
\end{thebibliography}
\end{document}